# Comparisons of the environmental dependence of galaxy properties between galaxies above and below M$^*$

Xin-Fa Deng   Ji-Zhou He   Xiao-Qing Wen   Xiao-Xun Tang

School of Science, Nanchang University, Jiangxi, China, 330031

**Abstract** From the Main galaxy sample of the Sloan Digital Sky Survey Data Release 6 (SDSS DR6), we construct two volume-limited samples above and below the value of M$^*$, to explore the difference of the environmental dependence of galaxy properties between galaxies above and below the value of M$^*$. We measure the local three-dimensional galaxy density in a comoving sphere with a radius of the distance to the 5th nearest galaxy for each galaxy, and compare basic properties of galaxies in the lowest density regime with those of galaxies in the densest regime. It is found that the galaxy luminosity strongly depend on local environments only for galaxies above M$^*$, but this dependence is very weak for galaxies below M$^*$. It is worth noting that g-r color, concentration index ci and galaxy morphologies strongly depend on local environments for all galaxies with different luminosities. This shows that M$^*$ is an characteristic parameter only for the environmental dependence of galaxy luminosity.

**Keywords**   Galaxy: fundamental parameters

## 1. Introduction

For a long time, correlations between galaxy properties and environment have been an important issue of Astrophysics (e.g., Blanton et al. 2003; Goto et al. 2003; Helsdon & Ponman 2003; Hogg et al. 2003, 2004; Treu et al. 2003; Balogh et al. 2004a, 2004b; Kauffmann  et al. 2004; Tanaka et al. 2004; Berlind et al. 2005; Blanton et al. 2005; Croton et al. 2005; Hoyle et al. 2005; Blanton & Berlind 2007; Deng et al. 2007a-b, 2008a-b; Park et al. 2007). Blanton et al. (2003) presented the dependence of seven galaxy properties on local galaxy density. Goto et al.(2003) found the expected trend of increasing early type fraction with increasing density. Deng et al. (2007a) showed that luminous galaxies exist preferentially in the densest regions of the universe (e.g., in groups), while faint galaxies are located preferentially in low density regions. It is widely accepted that luminous, red and early type galaxies exist preferentially in the densest regions of the universe, while faint, blue and late type galaxies are located preferentially in low density regions. Norberg et al. (2001) found that the clustering amplitude increases slowly with absolute magnitude for galaxies fainter than $M^*_{bJ}$-5 $\log_{10}$ h=-19.7 (Folkes et al.1999), but rises more strongly at higher luminosities. Thus, we believe that galaxies above and below the value of M$^*$ found for the overall Schechter fit to the galaxy luminosity function may have different environmental dependence of galaxy properties, and intend to explore this difference between them.

In the past, many authors measured the local galaxy density around each galaxy and investigated the dependence of galaxy properties on local density. But it is difficult to find a perfect statistical method and density measure for this issue. We note that in previous studies, different statistical methods and density measures were used, to get a convincing conclusion. For example, Balogh et al. (2004b) used two Gaussian distributions to fit the color distribution of galaxies, divided the sample into a blue and red population, and found that at fixed luminosity the mean color of blue galaxies or red galaxies is nearly independent of environment. Deng et al. (2008a) compared distributions of basic galaxy properties in the lowest density regime with those in the densest regime, and found that that the sample at low density has a higher proportion of faint, blue and late type galaxies and a lower proportion of luminous, red and early type galaxies than the sample at high density. In this study, like Deng et al. (2008a), we measure the density within the distance to the 5th nearest neighbor, and compare basic properties of galaxies in the lowest density regime with those of galaxies in the densest regime, to investigate the local environmental dependence of galaxy properties. This is a simple and effective method. Our paper is organized as follows. In section 2, we describe the data used. The environmental dependence of galaxy properties for galaxies above and below the value of $M^*$ are discussed in section 3. Our main results and conclusions are summarized in section 4.

In calculating the distance we used a cosmological model with a matter density $\Omega_0 = 0.3$, cosmological constant $\Omega_\Lambda = 0.7$, Hubble's constant $H_0 = 70 \text{km} \cdot \text{s}^{-1} \cdot \text{Mpc}^{-1}$.

## 2. Data

Many of survey properties of the SDSS were discussed in detail in the Early Data Release paper (Stoughton et al. 2002). Galaxy spectroscopic targets were selected by two algorithms. The Main galaxy sample (Strauss et al. 2002) comprises galaxies brighter than $r_{petro}$ < 17.77(r-band apparent Petrosian magnitude). The Luminous Red Galaxy (LRG) algorithm (Eisenstein et al. 2001) selects galaxies to $r_{petro}$ < 19.5 that are likely to be luminous early-types, based on the observed colors. We use the Main galaxy sample (Strauss et al. 2002). In our work, the data were downloaded from the Catalog Archive Server of SDSS Data Release 6 (Adelman-McCarthy et al. 2008) by the SDSS SQL Search (with SDSS flag: bestPrimtarget&64>0) with high-confidence redshifts (Zwarning $\neq 16$ and Zstatus $\neq 0$, 1 and redshift confidence level: zconf>0.95) (http://www.sdss.org/dr6/).

Schawinski et al. (2007) indicated that at z < 0.05, SDSS spectroscopy begins to be incomplete for bright galaxies. Thus, we construct a luminous volume-limited Main galaxy sample which contains 94954 galaxies at 0.05<z<0.089 with -22.40<$M_r$<-20.16. The absolute magnitude $M_r$ is calculated from the r-band apparent Petrosian magnitude, using a polynomial fit formula (Park et al. 2005a) for the mean K-correction within 0 < z < 0.3:
K(z) = 2.3537(z−0.1)$^2$+1.04423(z−0.1)−2.5log(1+0.1).

The lower luminosity limit $M_r = -20.16$ corresponds to 0.33 mag fainter than the value of $M^*$ found for the overall Schechter fit to the galaxy luminosity function (Ball et al. 2006). So, most galaxies in this sample are brighter than $M^*$. We also construct a volume-limited sample fainter than $M^*$, which is limited within the redshift range $0.02 \leq z \leq 0.0436$ and contains 21878 galaxies with the luminosity -20.0 $\leq M_r \leq$ -18.5.

## 3. Environmental dependence of galaxy properties for galaxies above and below $M^*$

Three-dimensional density estimator is severely biased by galaxy peculiar velocities, especially in dense region. We note that many authors measured the projected local density $\Sigma_5$ which is computed from the distance to the 5th nearest neighbor within a redshift slice $\pm 1000$ km s$^{-1}$ of each galaxy (e.g., Goto et al. 2003; Balogh et al. 2004a, 2004b). Although the projected local density $\Sigma_5$ is free from the influence from the radial velocity, it only is a projected quantity, and is influenced by projection effects. In fact, it is very difficult to find a density measure which does not introduce any biases. Maybe, we should try different density measures which may introduce different biases, to get a reliable conclusion about this issue.

Like Deng et al. (2008a), we compute the local three-dimensional galaxy density in a comoving sphere with a radius of the distance to the 5th nearest galaxy for each galaxy, and arrange galaxies in a density order from the smallest to the largest. For each sample, we select about 5% galaxies and construct two subsamples at both extremes of density according to the density.

$R_{50}$ and $R_{90}$ are the radii enclosing 50% and 90% of the Petrosian flux, respectively. In this study, we use the concentration index of r-band ci $= R_{90}/R_{50}$ as a structural parameter. Fig.1-3 show the luminosity, g-r color and concentration index ci distributions of two subsamples at both extremes of density for the faint and luminous volume-limited samples. As seen from Figure 1, in the luminous volume-limited sample, strong environmental dependence of the luminosity can be observed: the subsample at low density has a higher proportion of faint galaxies ($M_r \geq -20.83$) and a lower proportion of luminous galaxies ($M_r \leq -21.28$) than the one at high density, and the level of significance is at least $4\sigma$ in most bins. But in the faint volume-limited sample, we do not find significant environmental dependence of the luminosity. We also perform a Kolmogorov-Smirnov (KS) Test of the luminosity distributions of two subsamples at both extremes of density. The probability of the two distributions coming from the same parent distribution is listed in Table 1. In the faint volume-limited sample, this probability is 0.00401, which is much larger than the one in the luminous volume-limited sample. Our result is consistent with previous conclusions found by the correlation function (e.g., Norberg et al. 2001). This shows that $M^*$ is an important characteristic parameter for the environmental dependence of galaxy luminosity or changes of clustering parameters of galaxies with luminosity. So, we conclude preferentially that when investigating of the environmental dependence of galaxy luminosity, galaxy samples fainter than $M^*$ and ones brighter than $M^*$ should be studied respectively.

But in Fig2-3, significant difference does not found between the faint and luminous volume-limited samples. K–S Test also is in good agreement with this conclusion. K–S probability of g-r color and concentration index ci nearly is 0, which shows that two distributions of g-r color and concentration index ci at both extremes of density completely differ, and that for the faint and luminous volume-limited samples there is the existence of strong environmental dependence of g-r color and concentration index ci: red and high concentration galaxies exist preferentially in the densest regions of the universe, while blue and low concentration galaxies are located preferentially in low density regions, which is in good agreement with the conclusion obtained by Deng et al. (2008b). The environmental dependence of galaxy luminosity for galaxies above and below the value of $M^*$ is fairly different: luminous galaxies show strong environmental dependence of the luminosity, but the luminosity of faint galaxies is only weakly correlated with environment. For the environmental dependence of g-r color and concentration index ci, such a difference does not found between galaxies above and below the value of $M^*$. This shows that $M^*$ is an characteristic parameter only for the environmental dependence of galaxy luminosity.

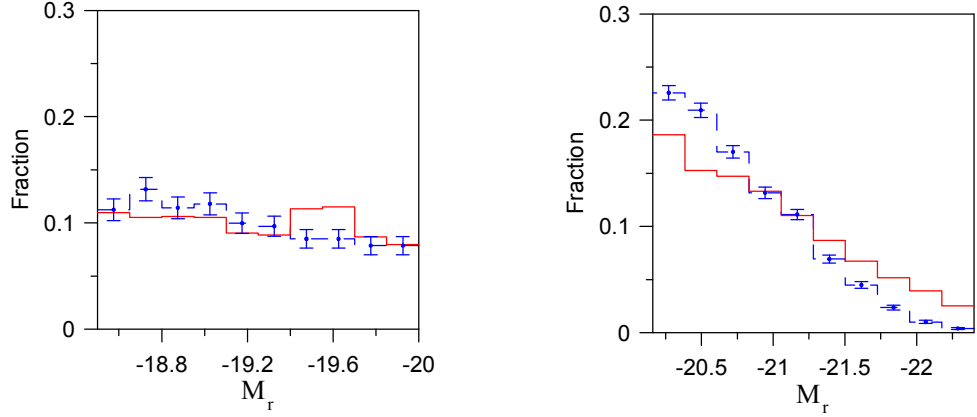

Fig.1 The luminosity distribution at both extremes of density for the faint (left panel) and luminous(right panel) volume-limited samples: red solid line for the subsample at high density, blue dashed line for the subsample at low density. The error bars of blue lines are 1 $\sigma$ Poissonian errors. Error-bars of red lines are omitted for clarity.

Bamford et al.(2009) investigated the relationships between galaxy morphology, color, environment and stellar mass, and found that at fixed stellar mass, color is highly sensitive to environment, while morphology displays much weaker environmental trends. In this study, the concentration index ci $= R_{90}/R_{50}$ is used to separate early-type (ci $\geq 2.86$) galaxies from late-type (ci <2.86) galaxies (Shimasaku et al. 2001, Nakamura et al. 2003). In fact, in the SDSS, it is nearly impossible to classify SDSS galaxies into morphological classes through direct inspection of the galaxy images as previous studies, due to a large number of galaxies. Whether a galaxy is termed 'early' or 'late' is fairly subjective. Many authors developed different methods or used other parameters, such as color, and star formation rate indicators as the morphology classification tool (e.g., Shimasaku et al. 2001; Strateva et al. 2001; Abraham et al. 2003; Park & Choi 2005b; Yamauchi et al. 2005; Sorrentino et al. 2006). But it is difficult to say which method or parameter is the best choice. Park & Choi (2005b) used the color versus color gradient space as the major morphology classification tool and the concentration index as an auxiliary parameter. But as seen from the upper panel of Fig.1 of Park & Choi (2005b), the concentration index is still a relatively good and simple parameter to be used to classify morphology of galaxies. We compute the early-type fraction of two subsamples at both extremes of density for each sample: for the faint volume-limited sample, 11.97% at low density and 21.39% at high density; for the luminous volume-limited sample, 24.83% at low density and 44.57% at high density. For the faint and luminous volume-limited samples, galaxy morphologies all strongly depend on local environments: galaxies in dense environments have predominantly early type morphologies, which were confirmed by many other studies. For example, Croton et al. (2005) showed that the population in voids is dominated by late types, in contrast, cluster regions have a relative excess of very bright early-type galaxies. This suggests that in dense environments there exist the transformation of galaxies from late to early types, which can be explained by some physical mechanisms, such as galaxy harassment (Moore et al. 1996), rampressure stripping (Gunn & Gott 1972) and galaxy-galaxy merging (Toomre & Toomre 1972).

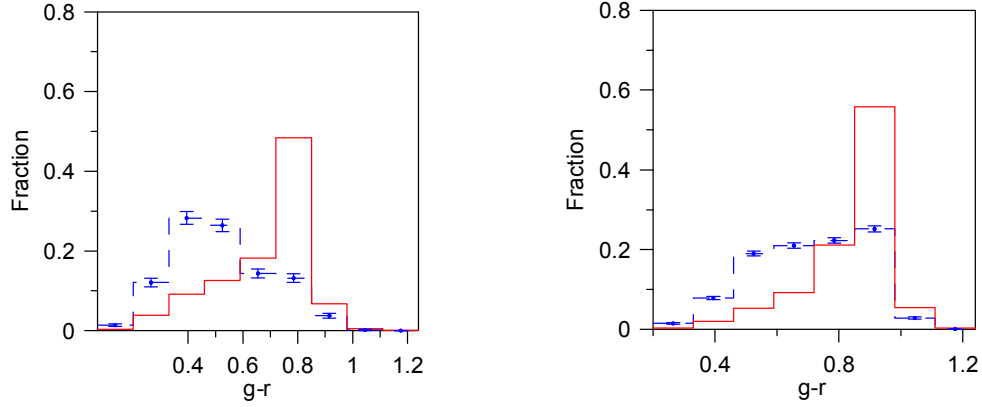

Fig.2 The g-r color distribution at both extremes of density for the faint (left panel) and luminous (right panel) volume-limited samples: red solid line for the subsample at high density, blue dashed line for the subsample at low density. The error bars of blue lines are 1 $\sigma$ Poissonian errors. Error-bars of red lines are omitted for clarity.

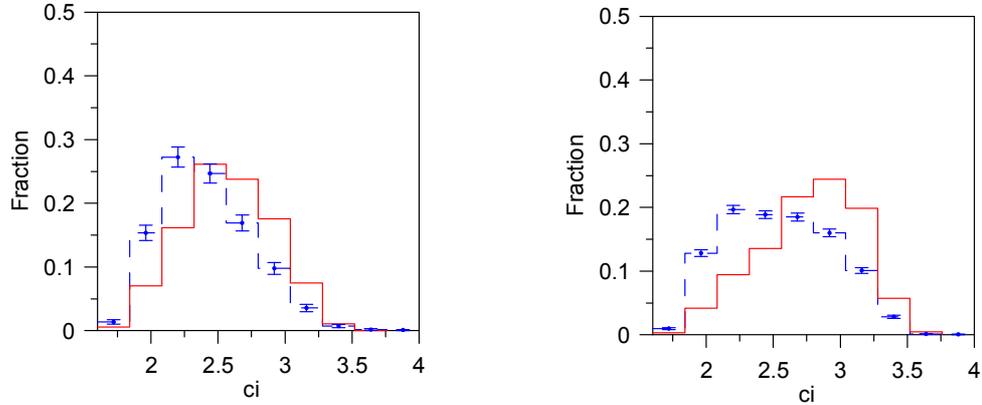

Fig.3 The ci distribution at both extremes of density for the faint (left panel) and luminous (right panel) volume-limited samples: red solid line for the subsample at high density, blue dashed line for the subsample at low density. The error bars of blue lines are 1 $\sigma$ Poissonian errors. Error-bars of red lines are omitted for clarity.

## 4. Summary

From the Main galaxy sample of SDSS DR6, we construct two volume-limited Main galaxy samples above and below the value of $M^*$, to explore the difference of the environmental dependence of galaxy properties between galaxies above and below the value of $M^*$. It is found that the galaxy luminosity strongly depend on local environments only for galaxies above $M^*$, but this dependence is very weak for galaxies below $M^*$. It is worth noting that g-r color, concentration index ci and galaxy morphologies strongly depend on local environments for all galaxies with different luminosity. This shows that $M^*$ is an characteristic parameter only for the environmental dependence of galaxy luminosity.

Table 1: K–S probabilities of the luminosity, g-r color and concentration index ci that two subsamples at both extremes of density are drawn from the same distribution.

| The sample | P(luminosity) | P(g-r color) | P(ci) |
|---|---|---|---|
| faint volume-limited sample | 0.00401 | 0 | 2.593e-24 |
| luminous volume-limited sample | 2.075e--33 | 0 | 0 |

**Acknowledgements**

We thank the anonymous referee for many useful comments and suggestions. Our study was supported by the National Natural Science Foundation of China (NSFC, Grant 10863002) and also supported by the Program for Innovative Research Team of Nanchang University.

Funding for the SDSS and SDSS-II has been provided by the Alfred P. Sloan Foundation, the Participating Institutions, the National Science Foundation, the US Department of Energy, the National Aeronautics and Space Administration, the Japanese Monbukagakusho, the Max Planck Society, and the Higher Education Funding Council for England. The SDSSWeb site is http://www.sdss.org.

The SDSS is managed by the Astrophysical Research Consortium for the Participating Institutions. The Participating Institutions are the American Museumof Natural History, Astrophysical Institute Potsdam, University of Basel, University of Cambridge, Case Western Reserve University, University of Chicago, Drexel University, Fermilab, the Institute for Advanced Study, the Japan Participation Group, Johns Hopkins University, the Joint Institute for Nuclear Astrophysics, the Kavli Institute for Particle Astrophysics and Cosmology, the Korean Scientist Group, the Chinese Academy of Sciences (LAMOST), Los Alamos National Laboratory, the Max Planck Institute for Astronomy (MPIA), the Max Planck Institute for Astrophysics (MPA), New Mexico State University, Ohio State University, University of Pittsburgh, University of Portsmouth, Princeton University, the US Naval Observatory, and the University of Washington.